# AI and Cultural Context: An Empirical Investigation of Large Language Models' Performance on Chinese Social Work Professional Standards

Zia Qi[1], Brian E. Perron[1], Miao Wang[2], Cao Fang[3], Sitao Chen[4], Bryan G. Victor[5]

[1]School of Social Work, University of Michigan

[2]Department of Social Work and Social Policy, Nankai University

[3]George Warren Brown School of Social Work, Washington University in St. Louis

[4]School of Social Work, University of Pittsburgh

[5]School of Social Work, Wayne State University

Correspondence concerning this article should be addressed to Zia Qi, University of Michigan, School of Social Work. 1080 South University Avenue, Ann Arbor, MI 48109. qizixuan@umich.edu

**Author Note**

Zia Qi https://orcid.org/0000-0002-8407-0465

Brian E. Perron https://orcid.org/0009-0008-4865-451X

Miao Wang https://orcid.org/0000-0002-1024-0513

Cao Fang https://orcid.org/0009-0002-2172-3235

Sitao Chen https://orcid.org/0009-0008-6037-6997

Bryan G. Victor https://orcid.org/0000-0002-2092-912X

**Disclosures**

We have no known conflict of interest to disclose. This article has been submitted for publication in the Journal of the Society for Social Work and Research. AI tools were used to assist with text editing and code generation, but all work is original and authored solely by the contributors.

2# Abstract

**Objective:** This study examines how well leading Chinese and Western large language models understand and apply Chinese social work principles, focusing on their foundational knowledge within a non-Western professional setting. We test whether the cultural context in the developing country influences model reasoning and accuracy.

**Method:** Using a published self-study version of the Chinese National Social Work Examination (160 questions) covering jurisprudence and applied knowledge, we administered three testing conditions to eight cloud-based large language models—four Chinese and four Western. We examined their responses following official guidelines and evaluated their explanations' reasoning quality.

**Results:** Seven of eight models surpassed the passing threshold of 60 points on both test sections, with only one model falling slightly below (59.5) in the applied knowledge section. Chinese models performed stronger on jurisprudence questions (median = 77.0 vs. 70.3 for Western models) but somewhat lower on applied knowledge (median = 65.5 vs. 67.0). Both Chinese and Western models showed cultural biases, particularly in scenarios involving gender equality and family dynamics. Models demonstrated a strong command of professional terminology but often failed to recognize culturally specific intervention techniques. Expert review revealed rates of valid reasoning among incorrect answers ranged from 16.4% to 45.0%.

**Conclusions:** Chinese and Western models demonstrate foundational knowledge of Chinese social work principles, though performance patterns vary by domains and question types. While Chinese models show advantages in regulatory content, Chinese and Western models struggle with culturally nuanced practice scenarios. These findings suggest that technical language ability does not guarantee cultural competence. The results contribute to informing responsible AI integration into cross-cultural social work practice.

**Keywords:** Large Language Models (LLMs), Chinese Social Work, Cross-Cultural Assessment, Professional Licensure, Artificial Intelligence in Social Work

# AI and Cultural Context: An Empirical Investigation of Large Language Models' Performance on Chinese Social Work Professional Standards

The field of social work stands at a critical juncture where the question is no longer *whether* to engage with AI technologies but *how* to do so thoughtfully, effectively, contextually, and ethically. While vibrant discussions explore AI's potential to advance research, education, and practice, equally important voices raise concerns about bias, data privacy, and environmental impact (e.g., Goldkind, Wolf, Glennon, Rios, & Nissen, 2023; National Association of Social Workers, n.d.; Patton, Landau, Mathiyazhagan, 2023; Reamer, 2004; Seniutis, Gružauskas, Lileikienė, & Navickas, 2024). As authors, we approach this intersection with the conviction that empirical evidence must enrich these philosophical debates.

An important first step is to understand how well these models comprehend and reason using the language of social work in different use cases. Recent research has revealed promising capabilities. For example, Victor, Kubiak, Angell, and Perron (2023) demonstrated that ChatGPT could understand the contents of the Association of Social Work Boards licensing examination and provide responses deemed even more appropriate than those offered by the test developers, suggesting the model possessed robust knowledge of Western social work principles. Luan and colleagues (Luan, Perron, Victor, Wan, Niu, & Xiao, 2024; Luan, Perron, & Zhang, 2024) extended this understanding by demonstrating ChatGPT's ability to process scientific literature about left-behind children in both English and Chinese, achieving excellent performance in summarizing and extracting information from academic abstracts. Further advancing our understanding, a recent study demonstrated the practical potential of connecting ChatGPT to an external knowledge base for answering specific questions about educational policies and procedures in a school of social work (Perron, Hiltz, Khang, & Savas, 2024). This approach effectively minimized hallucinations while providing contextually relevant responses.

While these findings are encouraging, a fundamental gap remains in our understanding of LLMs in social work – that is, we have not evaluated whether and to what extent leading LLMs can engage with social work content outside of Western contexts. As AI increasingly permeates global social work practice, we need to continue assessing model performance beyond Western contexts. China provides a particularly informative setting due to its distinctive cultural landscape, institutional arrangements, and rapidly evolving social work field, where professional standards, training, and policy guidance are developing at a remarkable pace. At the same time, China is emerging as a major hub of AI innovation, supported by substantial government and private sector investment, robust research infrastructures, and a growing talent pool of data scientists and engineers. Parallel developments in social work and AI create new opportunities for examining how LLMs may function in complex social service ecosystems.

This understanding is complicated by the Western-centric dominance of major AI companies, with OpenAI's ChatGPT commanding the most extensive user base by orders of magnitude. Indeed, ChatGPT has become so ubiquitous that it has undergone genericization in everyday language, with people using "ChatGPT" to refer to LLMs generally, much like "Google" as a verb for web searching (e.g., "Just google it."). This linguistic phenomenon reflects a more profound concern – the potential for Western AI tools to become de facto standards globally without sufficient examination of their cultural appropriateness or effectiveness in non-Western contexts.

Moreover, although cloud-based LLMs such as ChatGPT, Claude, and Gemini hold significant promise for advancing research and practice, they remain inaccessible to many regions due to regulatory and legal barriers, language and cultural complexities, and infrastructure limitations. Even when technically equipped to process non-Western languages and content, these models often struggle to capture deep cultural nuances and the professional frameworks that shape social work in diverse contexts. This cultural bias in LLMs has been observed in multiple studies. For instance, Tao, Vibert, Baker, and Kizilcec (2024) found that popular LLMs like GPT-4 and GPT-3.5 exhibit cultural values resembling English-speaking and Protestant European countries, demonstrating a clear Western bias. While recent findings suggest that ChatGPT can handle Chinese-language scientific abstracts on left-behind children (see Luan et al., 2024a, 2024b), surface-level linguistic capabilities do not equate to an understanding of the cultural, moral, and professional norms underlying social work practice in China. Such understanding often derives from a synthesis of knowledge sources that extend beyond textual information, including local training and credentialing systems, community values, and case-specific approaches. This aligns with findings from Cao et al. (2023), who demonstrated that LLMs struggle to accurately represent local cultural values even when prompted in different languages.

In this sense, the technical and geopolitical factors limiting the global knowledge exchange—such as the uneven distribution of AI technologies, the dominance of Western-based AI companies, and language or data-access disparities—reinforce a Western-centric model of AI development and evaluation. This dynamic limits the adaptability and cultural responsiveness of these tools and places non-Western regions at a disadvantage in shaping AI's future roles in social work. This concern is echoed by Johnson et al. (2022), who note that GPT-3 exhibits an "American accent" in its value conflicts, further emphasizing the Western bias in these models.

The challenge of cultural bias in LLMs extends beyond just language processing. As Tao et al. (2024) argue, the cultural bias in these models may inadvertently cause people to convey more Western-centric values in AI-assisted communication, potentially leading to misrepresentation and lack of cultural embeddedness in various contexts. This is particularly problematic in fields like social work, where cultural sensitivity and local context are essential for effective practice.

The combined effect of Western market dominance and the limited global accessibility of AI technologies risks creating a self-reinforcing cycle. Western models may become de facto standards, driving further Western-centric development and evaluation while potentially overlooking or misunderstanding the unique needs and contexts of social work practice in non-Western settings. This reality underscores the urgency of our research agenda to evaluate these models' capabilities within diverse cultural frameworks, particularly in settings like China, where social work practice operates under distinct philosophical and institutional paradigms. Rather than emphasizing bottom-up approaches, social work practices and priorities in China are closely aligned with government policies and socialist principles. This government-centric approach fundamentally distinguishes Chinese social work from its Western counterparts, making it a distinct case when considering the integration of Large Language Models into social work practice.

**Study Overview**

Our study advances the empirical understanding of AI capabilities in social work by conducting the first systematic cross-cultural benchmark of LLM performance on professional social work knowledge. Our analysis encompasses eight cloud-based models—evenly divided between major Western technology companies and leading Chinese AI firms. Moving beyond the methodological limitations of previous research that focused primarily on pass/fail metrics, we leverage the Chinese National Social Work Examination (CNSWE; National Social Worker Professional Compilation Group, 2024) to assess foundational professional knowledge. Two interconnected aims guide our investigation, specific to the Chinese social work context:

1. Evaluate social work knowledge across state-of-the-art LLMs: We assess the breadth and depth of social work knowledge embedded within current models, evaluating their understanding of core concepts, ethics, and policies within the Chinese context.
2. Assess model reasoning and knowledge application of LLMs: By analyzing confidence ratings and response patterns, we investigate how models reason through social work concepts, distinguishing between genuine understanding and pattern-matching behaviors. This analysis provides insights into the models' potential applications in professional contexts, considering their technical capabilities and cultural competencies.

**Methods**

To systematically evaluate the level of foundation knowledge of Chinese social work principles across LLMs, we required a comprehensive, culturally grounded assessment tool that could serve as a proxy measure for our target construct. The Chinese National Social Work Examination (CNSWE) emerged as an ideal instrument for two reasons. First, the standardized national licensure examination represents a consensus view of the essential knowledge required for social work practice in China, encompassing theoretical frameworks and practical applications within the Chinese context. Second, the exam's dual focus on jurisprudence and applied knowledge allows us to assess the models' understanding of Chinese social work's unique policy landscape and the culturally specific approaches to practice. Using performance on this examination as a proxy measure, we can empirically assess the depth of Chinese social work knowledge embedded within different LLMs' training.

**Chinese National Social Work Examination (CNSWE)**

The CNSWE is administered annually and is a mandatory requirement for obtaining professional certification and licensure in social work. It is a key assessment tool for evaluating the foundational knowledge and professional competence of social workers in China. It covers various topics, including social work theories, ethical principles, case management, community development, and specialty areas such as child welfare, mental health, and gerontology. This exam ensures that social work practitioners possess the necessary skills and understanding to effectively address the diverse and complex needs of individuals, families, and communities within the Chinese context. In the context of this study, the CNSWE provides a robust framework for evaluating the capabilities of LLMs in understanding and responding to culturally specific social work knowledge.

The examination system consists of assistant, intermediate, and advanced levels. For this study, we focus on the intermediate-level exam, which is composed of three major parts: a jurisprudence test covering social work policies and laws, an applied knowledge test covering a broad range of social work concepts, and scenario-based assessments that require candidates to demonstrate their ability to use social work principles in real-world situations (Zeng, Li, & Chen, 2019). Unlike the assistant level, the intermediate examination includes a critical jurisprudence component alongside the applied knowledge test and scenario-based assessments, making it suitable for evaluating LLMs' comprehension of both social work fundamentals and social work-related policies and laws, which is an important aspect of social work practice in the Chinese context. The advanced level exam, on the other hand, is entirely scenario-based and requires candidates to possess five years of work experience following the successful completion of the intermediate licensing exam, making it unsuitable for assessing LLMs' foundation knowledge in a standardized manner nor accurately reflecting the competencies of the majority of practicing professionals.

We selected the 2023 version of the intermediate level examination since this date coincides with our selected models' approximate training cutoff date. Thus, the models are likely to have been trained on existing social work knowledge but will have yet to be trained on the actual questions of the exam versions used in this study. While the complete examination consists of three parts (jurisprudence, applied knowledge, and scenario-based assessment), our study focuses on the jurisprudence and applied knowledge sections, as the scenario-based assessment requires a distinct research methodology beyond our current scope. While many exam questions are shared online, we use only questions from the exam guide materials published by the National Social Worker Professional Exam Question Compilation Group (2024a, 2024b).

Each exam section - jurisprudence test and applied knowledge test - follows identical structures in their question formats and scoring methods. The sections contain 80 questions each, combining two different question types. The first type consists of 60 single-select multiple-choice questions, where test-takers must choose one correct answer from four options. Each question is worth one point, contributing 60 points to the total score. The remaining 20 questions are in the select-all-that-apply (SATA) format using a partial credit scoring system. In STAT questions, test-takers must select two to four options from five choices. A fully correct answer earns two points, while a partially correct answer, containing only correct options, earns one-half point per correct choice. Selecting any incorrect option results in a score of zero for that question. The SATA section can contribute up to 40 points. Combined, the single-select and SATA questions create a possible composite score of 100 points for each section. According to the Ministry of Human Resources and Social Security (2022), the passing threshold for each test is 60% of the total possible score, which is 60 points. We maintain this scoring structure in the current study to ensure consistency with official examination standards.

We extracted all test questions and their associated response options from the examination guide to construct our test set and converted them to a structured data format. We maintained the original Chinese text while preserving all formatting and response option designations. Each question was assigned a unique identifier and tagged with the test type (i.e., jurisprudence or applied knowledge), question type (single-select or SATA), and content domain. This structured format enabled consistency across the experimental conditions and facilitated automated scoring. Examples of questions are provided in Appendix A in their original Chinese form and translated into English. The models were supplied with the question in Chinese in the actual tests.

**Selection of LLMs**

This study focused on frontier models, the most advanced LLMs developed by leading technology companies that rely on sophisticated architectures and cloud-based infrastructures. Selecting specific models presented unique challenges, as there are no comprehensive public databases tracking user statistics or standardized benchmark tests specific to social work applications. Additionally, many widely used models need more transparent usage metrics, making it difficult to rank their prevalence in professional settings definitively.

Given these constraints, we used a purposive selection strategy based on three critical criteria: (1) demonstrated prominence through media coverage and observed usage patterns; (2) sustained presence in technical discussions and benchmarking studies across multiple domains, indicating ongoing development and regular updates; and (3) availability of stable API access. This last criterion was particularly important as it enabled standardized testing procedures and eliminated the variability inherent in web-based interfaces, indicating the model's maturity and intended use in production environments rather than experimental applications.

The final selection includes eight models (see Table 1), evenly split between major Western technology companies and leading Chinese AI firms. This balanced representation allows for meaningful comparison while acknowledging the distinct technological and cultural contexts in which these models were developed. While other models exist, these selections represent a robust sample of current state-of-the-art LLMs with an established presence in professional settings. We acknowledge that model capabilities continue to evolve rapidly, and our findings should be interpreted as a snapshot of performance at the time of testing rather than a definitive ranking.

Table 1. List of Chinese and Western Large Language Models Used in Performance Testing

| Model Name | Region | Company |
| --- | --- | --- |
| Ernie-4.0-8k | Chinese | Baidu |
| Qwen-Max | Chinese | Alibaba |
| DeepSeek-2.5 | Chinese | DeepSeek AI |
| Moonshot-v1-8k | Chinese | Moonshot AI |
| Claude-3.5-Sonnet | Western | Anthropic |
| Gemini-1.5-Pro | Western | Google/Alphabet |
| ChatGPT-4o | Western | OpenAI |
| Mistral-Large | Western | Mistral AI |

**Test Conditions**

Our study evaluated LLM performance using three distinct testing conditions, each designed to assess different aspects of model capabilities in Chinese social work. These conditions were developed based on prior research methodologies (see Victor et al. 2024) while introducing novel elements specific to the Chinese social work context.

In Condition 1 (required response), we replicated traditional examination conditions, establishing our baseline for comparison. Models received explicit instructions regarding question type (single-select or SATA) and were directed to answer every question regardless of their certainty level. This baseline directly compared human performance metrics across jurisprudence and applied knowledge sections, comprising 60 single-select and 20 SATA questions. For each question, we asked the model to provide both an answer and a brief explanation of its reasoning, limiting explanations to approximately 150 words to match the exam guide's format. This dual-response approach allowed us to evaluate both answer accuracy and depth of understanding.

Condition 2 (selective response) modified the initial prompt to allow models to decline to answer questions about which they were uncertain. While maintaining similar procedures to Condition 1, models could skip questions they couldn't answer with reasonable confidence, though they still provided confidence ratings for attempted questions. This condition assessed the models' ability to recognize their knowledge limitations and capacity for self-assessment.

Condition 3 investigated potential examination design effects by testing for construct-irrelevant variance. Following methodologies from prior research (Victor et al., 2024; Albright & Thyer, 2010), we presented models with only answer options, withholding the actual questions. This approach tested whether the Chinese examination contained unintended patterns that could enable correct answers without understanding the content. Through simulation procedures using one million random samples, we calculated the expected guessing range based on the exam's combination of four-choice (single-select) and five-choice (SATA) question formats. This analysis established an expected random guessing range of 13.8% to 26.3%, with a theoretical mean of 19.9%.

**Data Management**

We developed an automated testing system to implement these conditions systematically using APIs to interact directly with the LLMs. An API is a standardized way to communicate with software programs that enable direct control and automation that is not possible through web-based interfaces (see Perron, Luan, Qi, Victor, & Goyal, 2024). This API-based approach provided fine-grained control over model parameters unavailable through web interfaces, allowing us to set the temperature to 0 to ensure consistent model responses. We developed a standardized prompting framework in Chinese that accommodated the specific requirements for each test condition, with English translations provided in Appendix A for research transparency.

Our analysis pipeline, built in Python (version 3.12; Python Software Foundation, 2024), leveraged these API capabilities to ensure efficient and consistent response processing. Using Pandas (version 2.0) for data manipulation and statistical analysis, we structured our workflow through Jupyter notebooks documenting each analytical step. The complete analysis workflow, including source code, configuration files, API implementation, and detailed documentation, is publicly available in our GitHub repository [https://github.com/ZiaQi/ai_and_cultural_context]. Final data visualizations were created using Tableau (v. 2022.4.0).

**Analytic Plan**

*Benchmark analysis*

Our analysis followed standard procedures for benchmarking LLM performance while incorporating methodological elements specific to professional licensing examination assessment. The analysis consisted of two major components: quantitative scoring and qualitative expert review of model responses. We computed composite scores for each quantitative component for each test following the official CNSWE scoring guidelines. This included calculating individual scores for single-select and SATA questions and overall composite scores. Consistent with standard practice in LLM benchmarking studies (e.g., Victor et al., 2024; Perron et al., 2024) and following the actual examination procedures, we focused on descriptive statistics rather than inferential statistical tests. This approach was chosen because professional licensing examinations use absolute performance thresholds rather than relative comparisons, the sample size of models (n=8) limits the meaningful application of statistical tests, the primary goal was to assess practical competency rather than establish statistical significance, and this approach maintains consistency with how scores are evaluated in the actual examination setting.

*Expert review of explanations*

Beyond quantitative performance metrics, we systematically analyzed models' explanations to assess their reasoning capabilities within Chinese social work contexts. This analysis involved two components: evaluating explanations for correct answers and analyzing reasoning patterns in incorrect responses.

For correct answers, we randomly sampled 30 responses from each mode. Each explanation was evaluated against the official examination guide using a binary classification: valid reasoning (either matching the guide or presenting sound alternative professional logic) or incorrect reasoning (correct answer but flawed professional rationale). For incorrect answers, we analyzed all 414 incorrect responses across models using a two-category classification system: unacceptable reasoning (indicating clear professional errors) and valid alternative reasoning (demonstrating sound professional judgment despite not matching official answers). All reviewers were bilingual social work professionals from mainland China with minimum MSW-level training, ensuring both linguistic and professional competency in evaluating responses within the Chinese social work context. The review process was facilitated through LabelBox, an online annotation platform that enabled systematic evaluation by multiple reviewers.

For evaluating inter-rater reliability in the analysis of model explanations, we used the Prevalence-Adjusted Bias-Adjusted Kappa (PABAK) (Byrt, Bishop, & Carlin, 1993). This adjusted version of the kappa statistic accounts for imbalances caused by differences in prevalence and bias in ratings between expert reviewers. Unlike the traditional kappa coefficient which can show poor reliability even when there is a high observed agreement between raters, PABAK provides an adjusted measure that depends solely on the observed proportion of agreement between raters. This is especially relevant to our study, where we found a high rate of valid reasoning with correct answers and a low rate with incorrect answers, skewing traditional reliability measures. Our reliability estimates indicate substantial agreement between raters for correct answers (.85) and moderate agreement for incorrect answers (.64), suggesting acceptable reliability overall (Cohen, 1992).

# Results

Our study evaluates the foundational knowledge of eight LLMs using the CNSWE across three test conditions. Our first test condition is the primary test, as it is consistent with the actual testing procedures. We then compare the scores on this first test condition with two variations to gain further insight into model performance.

**Condition 1: Required Response Format**

In this condition, models were instructed to answer every question. Seven of the eight models scored above the 60-point threshold on the jurisprudence and applied knowledge tests, with only Deepseek falling slightly below the threshold (59.5) on the applied knowledge section. The models exceeded chance performance by a wide margin. Across all models, performance was generally stronger on the jurisprudence than on the applied knowledge test (median scores = 71.0 and 66.0, respectively). The jurisprudence test showed wider performance variation, with scores ranging from 62.0 to 84.0, compared to a more restricted range on the applied knowledge test (59.5 to 71.0). When comparing Chinese and Western models, Chinese models scored slightly higher on the jurisprudence test (median = 77.0 vs. 70.3) but somewhat lower on the applied knowledge test (median = 65.5 vs. 67.0).

Figure 1 provides a breakdown of the scores for each model by test type. Six of eight LLMs performed better on the jurisprudence than on the applied knowledge test. On the jurisprudence test, the Chinese models, Ernie and Qwen, led by a substantial margin (84.0 and 83.0), exceeding their nearest competitors by ten percentage points. Claude-Sonnet (Western) tied for third with a score of 73.0, while Mistral (Western) had the lowest score of 62.0. Performance patterns differed on the applied knowledge test, where Mistral achieved the highest score (71.0), followed closely by Qwen (69.0) and Gemini (68.0). While Mistral showed manifest improvement from its jurisprudence score, gaining nine percentage points, Ernie demonstrated the most dramatic shift between tests, dropping 18 points from its jurisprudence score of 84.0 to 66.0 on the applied knowledge test.

**Figure 1.** Performance Comparison of Chinese and Western Large Language Models on the Jurisprudence and Applied Knowledge Test of the Chinese National Social Work Examination.

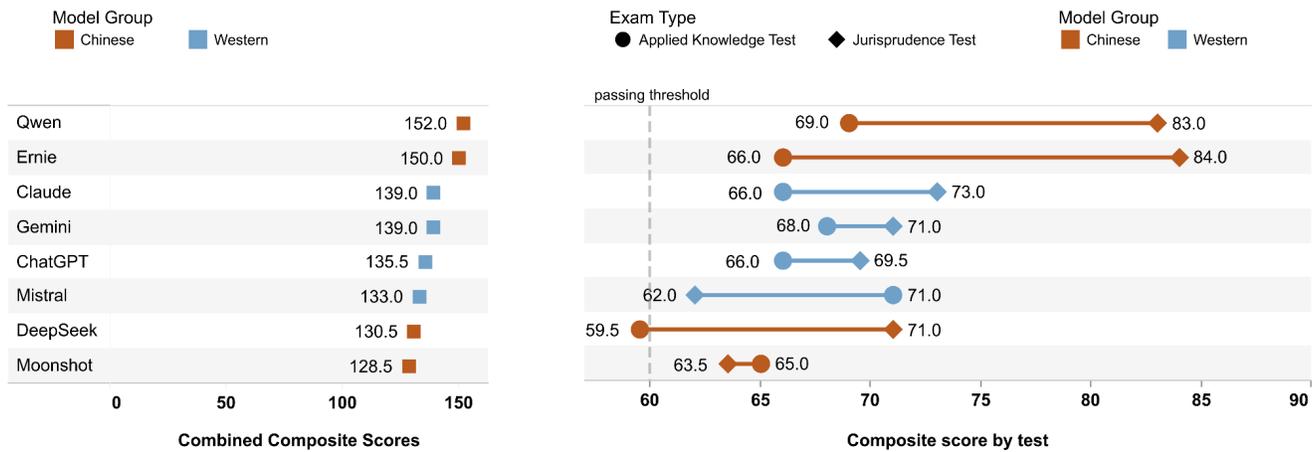

Combined Composite Scores are the sum of the Applied Knowledge Test (100 points) and the Jurisprudence Test (100 points; Total possible points = 200). Models tested: Ernie-4.0-8k, Qwen-Max, DeepSeek-2.5, Moonshot-v1-8k, Claude-3.5-Sonnet, Gemini-1.5-Pro, ChatGPT-4o, and Mistral-Large.

As demonstrated in Figure 2, performance patterns on the jurisprudence test revealed notable differences between single-select and SATA questions. The left panel shows raw scores, while the right panel displays normalized scores, dividing raw scores by the total points possible for each question type (60 points for single-select and 40 points for SATA). In both raw and normalized analyses, Chinese models demonstrated superior performance on single-select questions, with Ernie and Qwen achieving normalized scores above 85.0%, significantly outperforming their Western counterparts. However, these same models showed marked declines in performance on normalized SATA scores, dropping to the mid-70% range. In contrast, three Western models (Claude, ChatGPT, and Gemini) showed slightly improved performance on normalized SATA scores compared to their normalized single-select scores. However, their overall scores remained lower than the Chinese models. Mistral maintained relatively consistent normalized performance across both question types (62.0% for single-select and 61.0% for SATA). Across all models, normalized performance was generally stronger on single-select questions (median score = 74.5%) compared to SATA questions (median score = 65.0%), suggesting that the SATA format presented a particular challenge for the highest-performing Chinese models in assessing jurisprudence knowledge.

**Figure 2.** Performance Comparison Between Single-Select and Select-All-That-Apply (SATA) Questions on the Jurisprudence Test of the Chinese National Social Work Examination.

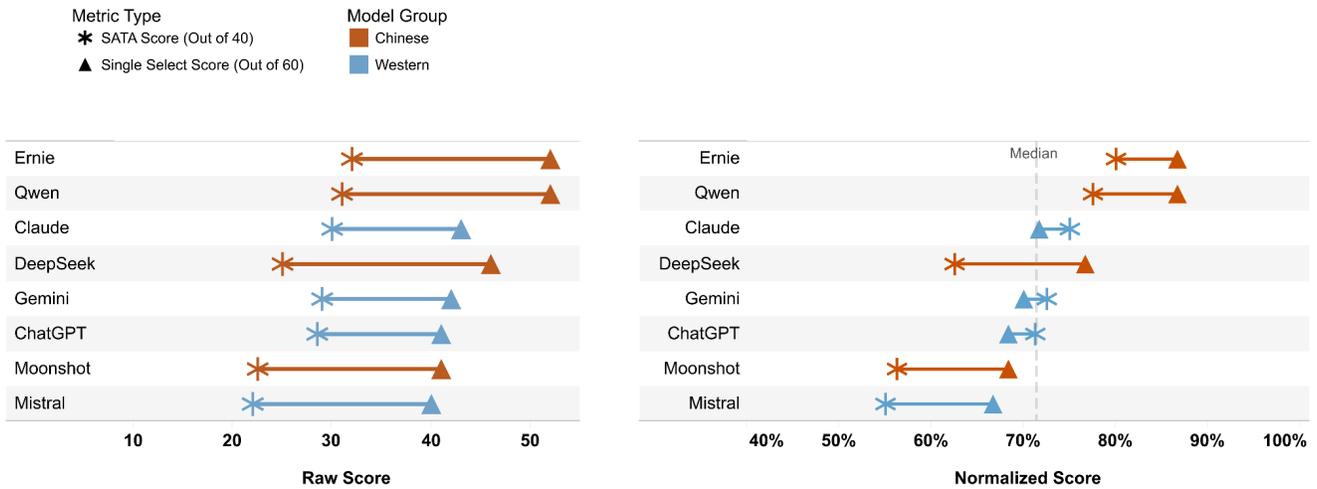

Models tested: Ernie-4.0-8k, Qwen-Max, DeepSeek-2.5, Moonshot-v1-8k, Claude-3.5-Sonnet, Gemini-1.5-Pro, ChatGPT-4o, and Mistral-Large. SATA = Select-all-that-apply.

Looking at Figure 3, the applied knowledge test performance patterns reveal a substantial contrast between single-select and SATA questions across all models. The raw score analysis shows models generally achieving higher performance on single-select questions than SATA, with most models scoring between 40-45 points (out of 60 possible) on single-select questions but only 20-25 points (out of 40 possible) on SATA questions. This performance gap becomes even more pronounced when examining normalized scores to account for the different point totals. In the normalized analysis, most models achieved scores between 65-75% on single-select questions but dropped to the 50-60% range on SATA questions, representing a consistent 15-20 percentage point decline in performance. This pattern held across Chinese and Western models, suggesting that the SATA format presented a universal challenge in assessing applied knowledge, regardless of the model's origin. This finding contrasts notably with the jurisprudence test results, where some Western models showed improved performance on SATA questions compared to single-select questions.

**Figure 3.** Performance Comparison Between Single-Select and Select-All-That-Apply Questions on the Applied Knowledge Test of the Chinese National Social Work Examination.

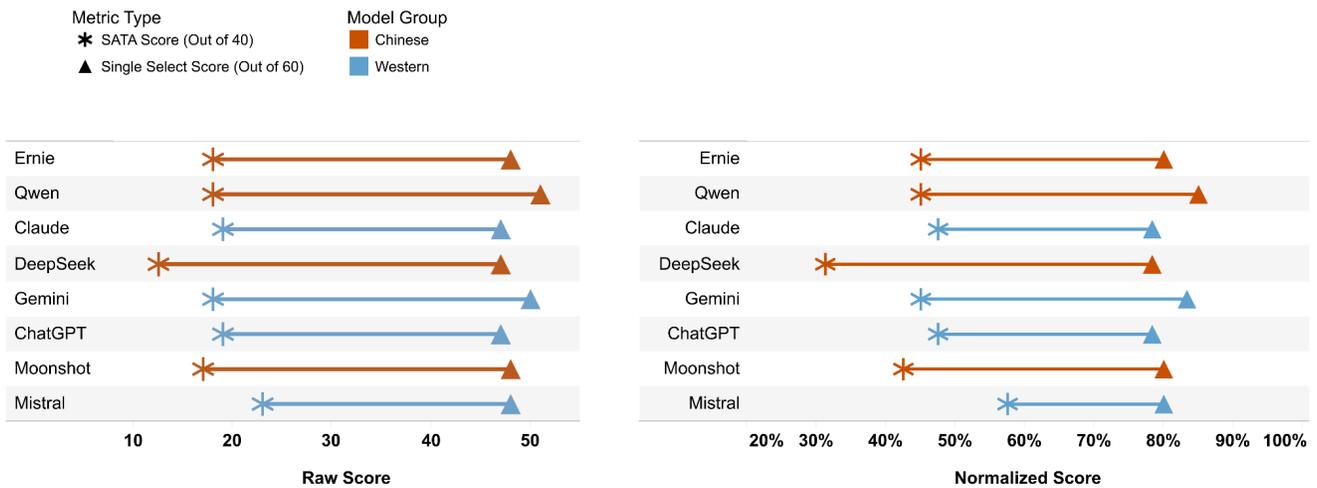

Models tested: Ernie-4.0-8k, Qwen-Max, DeepSeek-2.5, Moonshot-v1-8k, Claude-3.5-Sonnet, Gemini-1.5-Pro, ChatGPT-4o, and Mistral-Large. SATA = Select-all-that-apply.

**Condition 2: Option to Skip Format**

As shown in Table 2, when given the option to skip questions they were uncertain about on the jurisprudence test, five models exercised this option, while three (Deepseek, Claude, and Gemini) attempted all questions. ChatGPT and Mistral demonstrated the most conservative approach, each skipping four questions. Moonshot skipped two questions, while Qwen and Ernie each skipped one question despite being the top performers. The option to skip questions produced minor improvements in adjusted scores (i.e., percent correct), while performance patterns largely mirrored those seen in Condition 1. ChatGPT's accuracy improved from 61.2% to 64.5% when accounting for skipped questions, and Mistral's score increased from 55.0% to 57.9%. Qwen moved up one position to claim the top spot, while Ernie dropped one position to second place. The bottom three models (ChatGPT, Moonshot, and Mistral) maintained their relative positions from Condition 1, suggesting that skipping questions did not significantly impact their comparative standing on the jurisprudence section.

In contrast to the jurisprudence test, models showed different patterns of question-skipping behavior on the applied knowledge test (see Table 3). Only three models skipped questions. Specifically, ChatGPT and Mistral each skipped one question, while Qwen skipped one question despite its high performance. The option to skip questions produced minor improvements in adjusted scores (i.e., percent correct), while performance patterns showed notable differences from Condition 1. Ernie achieved the highest score with 71.2%, showing no skipped questions, while Qwen's score improved slightly from 68.8% to 69.6% when accounting for its skipped question. Ernie moved up three positions to claim the top spot, and Moonshot showed substantial improvement by moving up four positions to tie for third place with Claude and Gemini, each scoring 66.2%. The most dramatic shift occurred with Mistral, which dropped six positions despite skipping a question, while ChatGPT dropped three positions, ending with adjusted scores of 64.6%.

**Table 2.** Performance Comparison of Chinese and Western Large Language Models on the Jurisprudence Sections of the Chinese Social Work Professional Level Examination With an Option to Skip.

| Condition 2 Rank | Model | Total questions skipped | Total percent correct-w/Skip | Total percent correct w/o Skip | Change in Rank (compared to Condition 1) |
|---|---|---|---|---|---|
| 1 | Qwen | 1 | 81.0% | 80.0% | +1 |
| 2 | Ernie | 1 | 78.5% | 77.5% | -1 |
| 3 | Deepseek | 0 | 70.0% | 70.0% | +1 |
| 4 | Claude | 0 | 66.2% | 66.2% | -1 |
| 5 | Gemini | 0 | 65.0% | 65.0% | -1 |
| 6 | ChatGPT | 4 | 64.5% | 61.2% | 0 |
| 7 | Moonshot | 2 | 57.7% | 56.2% | 0 |
| 8 | Mistral | 4 | 57.9% | 55.0% | 0 |

Models are ranked based on total percent correct w/o skip in condition 2. Changes indicate the difference between Condition 2 and Condition 1 scores. Rankings are based on Condition 2 performance. Models tested: Ernie-4.0-8k, Qwen-Max, DeepSeek-2.5, Moonshot-v1-8k, Claude-3.5-Sonnet, Gemini-1.5-Pro, ChatGPT-4o, and Mistral-Large.

Table 3. Performance Comparison of Chinese and Western Large Language Models on the Applied Knowledge Sections of the Chinese Social Work Professional Level Examination

| Condition 2 Rank | Model | Total questions skipped | Total percent correct-w/Skip | Total percent correct w/o Skip | Change in Rank (compared to Condition 1) |
|---|---|---|---|---|---|
| 1 | Ernie | 0 | 71.2% | 71.2% | +3 |
| 2 | Qwen | 1 | 69.6% | 68.8% | 0 |
| = 3 | Moonshot | 0 | 66.2% | 66.2% | +4 |
| = 3 | Claude | 0 | 66.2% | 66.2% | +1 |
| = 3 | Gemini | 0 | 66.2% | 66.2% | 0 |
| 6 | Deepseek | 0 | 65.0% | 65.0% | +2 |
| = 7 | Mistral | 1 | 64.6% | 63.8% | -6 |
| = 7 | ChatGPT | 1 | 64.6% | 63.8% | -3 |

With an Option to Skip. Models are ranked based on total percent correct w/o skip in condition 2. Changes indicate the difference between Condition 2 and Condition 1 scores. Rankings are based on Condition 2 performance. Models tested: Ernie-4.0-8k, Qwen-Max, DeepSeek-2.5, Moonshot-v1-8k, Claude-3.5-Sonnet, Gemini-1.5-Pro, ChatGPT-4o, and Mistral-Large.

**Condition 3: Testing for Study Design Effects**

Condition 3 evaluated potential study design effects by presenting models with answer choices without their corresponding question stems. This approach tests whether models can correctly guess answers without seeing the actual questions. The expected range for random guessing was between 13.8% and 26.3%, with a theoretical mean of 19.9%. In our testing, all models fell outside this expected guessing range. The top-performing models were Claude and Qwen, both achieving 37.5%, followed closely by Ernie at 36.9%. While these scores exceeded chance levels, they were significantly lower than ChatGPT's 73.3% correct guessing rate observed by Victor et al. (2024) using the same methodology on the ASWB exam.

**Expert Review of Model Explanations**

Our expert review process revealed several key findings that illuminate how models engage with Chinese social work knowledge, each with important implications for understanding AI capabilities in cross-cultural professional contexts. First, our systematic review revealed a noteworthy pattern in reasoning validity that challenges conventional assessment approaches. For correct answers, as shown in Table 4, analysis of 30 randomly sampled responses per model showed expectedly high rates of valid reasoning (83.3% to 100%). More surprisingly, even for incorrect answers (414 total across models), models demonstrated substantial rates of valid reasoning, ranging from 16.4% to 45.0% (see Table 5). This suggests that a significant portion of "incorrect" answers may represent valid alternative approaches rather than failures of professional understanding. Ernie demonstrated this most frequently (45.0% of incorrect answers), followed by Gemini (37.5%) and Claude (36.4%). These cases often involved scenarios where multiple approaches could be professionally justified, though they didn't align with standardized testing requirements. For example, in questions about professional interventions, models sometimes provided theoretically sound alternatives that reflected international social work principles but didn't match the specific approaches emphasized in Chinese practice.

**Table 4.** Expert Review Classifications of Correct Answer Explanations by Model.

| Model | Valid Reasoning % (N)* | Invalid Reasoning % (N) |
|---|---|---|
| Chinese Models | | |
| Qwen | 96.7% (29) | 3.3% (1) |
| Moonshot | 96.7% (29) | 3.3% (1) |
| Ernie | 86.7% (26) | 13.3% (4) |
| DeepSeek | 83.3% (25) | 16.7% (5) |
| Western Models | | |
| Claude | 100% (30) | 0% (0) |
| Mistral | 96.7% (29) | 3.3% (1) |
| Gemini | 96.7% (29) | 3.3% (1) |
| ChatGPT | 90.0% (27) | 10.0% (3) |

*A random sample of 30 correct responses was analyzed for each model. Models tested: Ernie-4.0-8k, Qwen-Max, DeepSeek-2.5, Moonshot-v1-8k, Claude-3.5-Sonnet, Gemini-1.5-Pro, ChatGPT-4o, and Mistral-Large.

**Table 5.** Expert Review Classifications of Incorrect Answer Explanations by Model.

| Model | Total N | Valid Reasoning % (N) | Invalid Reasoning % (N) |
|---|---|---|---|
| Chinese Models | | | |
| Qwen | 38 | 34.2% (13) | 65.8% (25) |
| Ernie | 40 | 45.0% (18) | 55.0% (22) |
| DeepSeek | 57 | 31.6% (18) | 68.4% (39) |
| Moonshot | 61 | 16.4% (10) | 83.6% (51) |
| Western Models | | | |
| Claude | 55 | 36.4% (20) | 63.6% (35) |
| Gemini | 56 | 37.5% (21) | 62.5% (35) |
| ChatGPT | 57 | 28.1% (16) | 71.9% (41) |
| Mistral | 63 | 30.2% (19) | 69.8% (44) |

Models tested: Ernie-4.0-8k, Qwen-Max, DeepSeek-2.5, Moonshot-v1-8k, Claude-3.5-Sonnet, Gemini-1.5-Pro, ChatGPT-4o, and Mistral-Large.

Among the incorrect answers, Chinese models showed higher rates of valid reasoning for jurisprudence-related questions (median = 42.1% vs 33.2% for Western models) but similar rates to Western models on applied knowledge questions (median = 31.5% vs 32.8%). This pattern raises important questions about how we assess professional knowledge in standardized testing contexts, a theme we explore further in our discussion.

The review revealed nuanced patterns in how models engaged with professional terminology. While models demonstrated a strong command of formal bureaucratic language, including specialized terms like "文件精神" (spirit of the document), this linguistic competence did not consistently translate to professional understanding. This was particularly evident in responses to questions involving specialized social work concepts like "非评判" (non-judgmental) and "诊断" (diagnosis), where models sometimes conflated similar terms despite their distinct professional meanings. Moreover, we observed confusion around the term "社工" which can refer to social work, community work, or society work in different contexts, leading to inconsistent applications of professional knowledge.

Our analysis also identified systematic patterns in how models processed complex professional content. In questions involving mathematical computation, even when models correctly cited relevant policies, they provided inconsistent numerical answers. For instance, in questions about counseling room ratios in schools, models referenced the correct regulations but failed to accurately compute the required proportions. This pattern persisted even when numbers were written in Chinese characters rather than numerals. Models also showed particular difficulty with questions requiring sustained logical consistency, especially in SATA questions. This was notably evident in inheritance rights scenarios where models needed to track relationships across multiple generations. The challenge appeared more pronounced in jurisprudence questions containing multiple conditional statements or complex legal terms. Additionally, models often struggled to expand formal terminology into practical applications. When interpreting "企事业单位" (enterprises and public institutions), models tended toward overly literal interpretations rather than understanding the term's broader social context. Similarly, in group work scenarios, models often recognized basic intervention structures but missed cultural-specific elements of Chinese social work practice.

**Discussion**

Our study evaluated how effectively Chinese and Western LLMs understand and reason about social work principles in the Chinese context. While both types of models demonstrated sufficient competency to pass examination standards, their performance patterns varied systematically across different types of content. ChatGPT's relatively modest performance, despite its market leadership, highlights a notable disconnect between general capabilities and specialized cultural-professional knowledge, raising important questions about how we evaluate and deploy AI systems in professional settings.

The pattern of valid reasoning across models reveals a more nuanced picture than traditional testing metrics suggest. Models frequently provided valid alternative explanations even for technically incorrect answers, demonstrating meaningful engagement with professional concepts despite not matching standardized answers. This finding challenges conventional assessment methodologies and suggests that binary correct/incorrect scoring may inadequately capture these systems' nuanced understanding of professional concepts. The tension between standardized assessment and legitimate alternative reasoning patterns echoes ongoing debates about the limitations of multiple-choice testing in evaluating practical competence (Victor et al., 2024).

Unlike previous research with the ASWB examination (Victor et al., 2024), models showed a reduced ability to identify patterns in exam questions without accessing question stems, suggesting the Chinese social work examination may better minimize construct-irrelevant variance. This observation, combined with the high rate of valid alternative reasoning for incorrect answers, indicates a need to reevaluate how we assess professional competency through standardized testing.

These findings raise critical questions about the relationship between language processing capabilities and domain-specific knowledge in AI systems, particularly in cross-cultural professional contexts. They also challenge assumptions about the advantages of locally developed AI models in handling culturally specific content. Our analysis points to several key areas for examination: cultural competency in AI systems, the distinction between language capability and professional knowledge, the handling of professional standards across cultural contexts, current technological limitations, and implications for developing AI tools for global social work practice.

The following discussion explores these themes through our empirical findings, considering their implications for social work education, practice, and the development of culturally responsive AI tools.

**Cultural Competency and Language Processing**

Our findings reveal a complex relationship between language processing and cultural understanding in AI systems. First, the models' competence varies in handling cultural and policy content. While both Chinese and Western models demonstrated strong Chinese language competence, Chinese models excelled specifically in jurisprudence questions but showed no advantage in applied knowledge scenarios. This pattern challenges assumptions about locally developed AI models' superiority in handling cultural content. Despite their primary Chinese language training, Chinese models' advantage was limited to formal legal and policy content, likely reflecting the prevalence of official documents and regulatory materials in their training data.

Moreover, our study reveals a notable distinction in the models' ability to recognize cultural-specific professional techniques, particularly identifying group summarization methods. For example, when analyzing a group work scenario where a social worker summarized collective insights, "通过讨论，大家认识到自身和周边的力量来源，包括别人的信任和鼓励，家人的爱和朋友的陪伴" ("Through discussion, everyone came to realize the sources of strength in themselves and their surroundings, including others' trust and encouragement, family love, and companionship of friends"), Claude was the only model that correctly identified this as a professional summarization technique - a specialized skill in Chinese social work group practice. While other models demonstrated general language comprehension, they consistently failed to recognize the professional significance of this summarization approach, suggesting limitations in their understanding of culture-specific social work interventions. This finding highlights the importance of considering both professional practice competencies and linguistic and cultural competencies when evaluating model performance in specialized fields like social work.

Additionally, the models' performance suggests that command of regulatory language does not equate to deeper cultural understanding. This disparity appears linked to training data composition, where formal documents are well-represented but nuanced case studies and practical interventions are less common. This finding highlights the critical distinction between technical language processing and genuine cultural-professional competence, raising important questions about how training data availability shapes model performance in specialized domains. For example, in a question about identifying leadership qualities in a community handicraft group, ChatGPT uniquely identified "loving putting self in the spotlight" as a desirable leadership trait, reflecting a Western cultural perspective that contrasts sharply with Chinese social values that discourage individual prominence. Such examples highlight how technical language proficiency does not necessarily translate to cultural competency in social work practice.

More concerning, models perpetuated cultural biases despite clear professional and legal frameworks. This was evident in inheritance rights scenarios where both sons and daughters had equal legal standing under Chinese law, several models across both Chinese and Western origins exhibited patriarchal biases by favoring sons over daughters in their explanations despite explicit legal provisions for equal rights. These findings suggest that while models can process Chinese language effectively at a semantic level, they may inadvertently strengthen societal biases learned from training data, raising important considerations for their use in professional settings where equity and social justice are paramount.

**Technical Language vs. Professional Understanding**

Our analysis revealed an important distinction between models' ability to process Chinese language and their grasp of professional social work concepts. While models demonstrated a strong command of Chinese syntax and vocabulary, they often struggled to correctly understand professional terminologies in context. This was particularly evident in their explanations of incorrect answers, where models frequently showed valid linguistic comprehension but failed to apply proper social work principles.

For instance, Moonshot (a Chinese model) demonstrated this disconnect by citing sophisticated Chinese bureaucratic terminology like "文件精神" (spirit of the official document) in a jurisprudence question without a proper understanding of its meaning - language typically reserved for Communist Party and government meetings, which also coupled with the lowest rate of valid reasoning (16.4%) for incorrect responses among all models. This disparity was even more striking given that Moonshot performed worse than most Western models on jurisprudence questions. This suggests that a facility with a formal Chinese language does not necessarily indicate deeper professional understanding.

**Technical Limitations and Performance Patterns**

Our study revealed two major limitations in how these AI models handle complex social work questions. Both limitations stem from how these models work: instead of truly understanding and reasoning through problems as humans do, they predict what words should come next based on patterns they have seen before.

The first limitation became clear when models needed to solve math problems. Consider this example from our test: We asked models to calculate a tax deduction for someone who donated 60,000 yuan while having a taxable income of 200,000 yuan, using a 30% threshold. While the models could correctly quote the relevant tax laws, they gave wildly different answers to this straightforward calculation. This happened because - instead of doing math, models were merely predicting what numbers typically appear in discussions about tax calculations.

The second limitation showed up when models needed to follow complex logical steps. For example, when given a question about inheritance rights involving multiple family members across different generations, none of our eight tested models could correctly figure out who should inherit what. This occurred because the models were unable to maintain logical consistency throughout the multiple steps of reasoning; they were simply predicting what typically follows in similar discussions about inheritance.

**Supporting, not Replacing Decision-Making**

Our findings reveal important nuances about how LLMs can support, not replace, social work practice. While these models demonstrate strong capabilities in processing professional language and concepts, their performance on foundational knowledge assessments suggests their greatest potential lies in helping practitioners access and apply professional knowledge. Retrieval augmented generation (RAG) systems offer a particularly promising direction. By connecting LLMs' language processing capabilities with expert-curated knowledge bases, we could develop tools that deliver relevant information to practitioners at the point of need (see Perron et al. 2024b). For example, a RAG system could help a practitioner quickly access specific intervention strategies, policy guidelines, or evidence-based practices relevant to their current case, while preserving professional judgment about how to apply this information.

However, realizing this potential requires careful attention to human-computer interaction design. Future research should focus on understanding how practitioners seek and use information in their daily work and how AI-powered knowledge systems can best support these existing workflows. This includes investigating questions like: How should complex professional knowledge be organized and presented? What interface designs best support practitioners in critically evaluating and applying AI-retrieved information? How can we ensure these systems enhance rather than disrupt the therapeutic relationship?

**Strengths and Limitations**

Our study makes several important contributions to understanding how LLMs perform in non-Western professional contexts. Using a standardized national licensure examination, we provide the first systematic evaluation of Chinese social work knowledge across both Chinese and Western models. The dual focus on jurisprudence and applied knowledge sections allowed us to distinguish between cultural-specific and universal content understanding. Furthermore, our expert review process, conducted by bilingual social work professionals, provided valuable insights into how models reason about professional concepts across cultural contexts.

However, our methodology faces important limitations regarding the temporal relationship between model training and test content. Using the 2023 version of the Chinese National Social Work Examination, which coincides with model training cutoff dates, creates uncertainty about whether models were inadvertently exposed to exam questions during training. While we took steps to use only exam guide materials, we cannot definitively determine if performance reflects true knowledge or pattern matching from training data.

A significant limitation is our reliance on a professional licensing examination as the primary assessment tool. While the CNSWE provides a standardized measure of basic professional knowledge, it may not fully capture the nuanced understanding required for real-world social work practice. The multiple-choice format, even with expert review of explanations, cannot assess important aspects of professional competence, such as clinical judgment, cultural sensitivity, and ethical decision-making in complex scenarios. Further benchmarks incorporating case studies, open-ended responses, and practical assessments would provide a more comprehensive evaluation of model capabilities in social work contexts.

Our study has limitations in its binary classification of "Chinese" versus "Western" models, which oversimplifies the complex cultural influences on model development and performance. While we identified important architectural limitations in mathematical computation and logical reasoning, these findings represent only a snapshot in time given the rapid evolution of AI capabilities. During our study period, companies like DeepSeek, OpenAI, and Alibaba released new models with enhanced multi-step reasoning abilities, demonstrating the field's swift progress. This rapid advancement, coupled with the dynamic nature of Chinese social work practice, suggests the need for ongoing evaluation using diverse assessment approaches to fully understand these evolving technologies' capabilities and limitations.

**Conclusion**

Our analysis reveals the promise and limitations of current LLMs in cross-cultural social work contexts. While Chinese and Western models demonstrated sufficient competency to pass examination standards, their performance exposed notable limitations in handling nuanced content. The emergence of advanced reasoning models like DeepThink suggests promising directions forward, particularly in addressing the sequential prediction limitations of traditional LLMs. However, even these more sophisticated models display inconsistencies that reinforce the continued importance of human oversight in professional applications. By positioning AI as a knowledge delivery tool rather than an autonomous practitioner, we can develop systems that genuinely enhance professional practice while preserving the essential human elements of social work. Our study demonstrates these models' facility with foundational social work knowledge; the next step is leveraging this capability to create thoughtfully designed support systems that help practitioners better serve their clients.

# Appendix A: Example Questions

**Jurisprudence Test- Single Select Question**

根据《中华人民共和民国法典》, 王某立遗嘱时, 下列人员可以作为遗嘱见证人的是( )。

A. 赵某, 限制民事行为能力人, 不是继承人
B. 钱某, 完全民事行为能力人, 是受遗赠人
C. 孙某, 完全民事行为能力人, 是王某的债权人
D. 李某, 完全民事行为能力人, 是王某的主治医生, 无利害关系

English Translation:
According to the Civil Code of the People's Republic of China, which of the following persons can act as a witness to Wang's will?

A. Zhao, a person with limited civil capacity, who is not Wang's heir
B. Qian, a person with full civil capacity, who is a beneficiary
C. Sun, a person with full civil capacity, who is Wang's creditor
D. Li, a person with full civil capacity, who is Wang's attending physician with no stake in the matter

**Jurisprudence Test - Select All That Apply**

根据《中华人民共和未国成年人保护法》, 对临时监护的未成年人, 民政部门可以采取的安置方式包括( )。

A. 委托亲属抚养
B. 委托家庭寄养
C. 交由符合条件的申请人收养
D. 交由儿童福利机构进行抚养
E. 交由未成年人救助保护机构进行收留

English Translation:
According to the Minor Protection Law of the People's Republic of China, which of the following placement options can the Civil Affairs Department use for children under temporary guardianship? ( )

A. Place the child with relatives
B. Place the child with a foster family
C. Allow adoption by qualified applicants
D. Place the child in a welfare institution
E. Place the child in a youth protection facility

**Applied Knowledge Test - Single Select Question**

初中生小美的父母离异, 父亲因诈骗入狱, 她跟爷爷奶奶一起生活。小美经常听到周围邻居议论自己家的事情, 她也因此感到低人一等, 认为自己没有什么优点, 很自卑。小美不想让同学知道自己的情况, 与同学关系疏远, 总是独来独往。班主任老师观察到小美的情况后, 将其转介给学校社会工作者。根据增能理论, 社会工作者的下列做法, 最能体现个人层面增能的是( )。

A. 消除邻里对小美一家人的偏见
B. 提升小美应对其他人歧视的能力
C. 为小美一家争取社区系统支持
D. 邀请小美参加社区儿童支持小组

English Translation:

Xiaomei is a middle school student whose parents are divorced and whose father is in prison for fraud. She lives with her grandparents. Xiaomei often overhears neighbors gossiping about her family situation, which makes her feel inferior. She believes she has no good qualities and suffers from low self-esteem. Not wanting her classmates to know about her situation, she keeps her distance from them and tends to be alone. After observing this, her school teacher referred her to the school social worker. According to empowerment theory, which of the following actions by the social worker best reflects individual-level empowerment? ( )

A. Eliminate neighbors' prejudice against Xiaomei's family

B. Enhance Xiaomei's ability to cope with discrimination from others

C. Seek community system support for Xiaomei's family

D. Invite Xiaomei to join a community children's support group

**Applied Knowledge Test - Select All That Apply**

社会工作者小周在一次个案面谈中得知，服务对象小李已成功戒毒，但在吸毒期间染上了艾滋病。小李因为害怕失去妻子，要求小周一定为他保密。妻子则经常向小周抱怨小李行为怪异，对自己感情冷淡，怀疑他对婚姻不忠，并希望通过怀孕来保全自己的婚姻和家庭。根据社会工作专业伦理，小周宜采取的以下做法有（）。

A. 将小李的病情直接告知其妻子，请她多加关注
B. 为小李疏导情绪，减轻精神压力积极面对问题
C. 征得小李同意后，为他介绍病友自助互助小组
D. 将小李的全部情况在机构个案报告会议中讨论
E. 与小李的妻子探讨该如何维系他们的婚姻关系

English Translation:
Social worker Zhou learns that his client Li has successfully recovered from drug addiction but contracted HIV during his period of drug use. Li begs Zhou to keep this information confidential, fearing his wife will leave him. Meanwhile, Li's wife frequently confides in Zhou about her husband's strange behaviors and emotional distance. She suspects he's being unfaithful and is considering getting pregnant to save their marriage. According to social work ethics, which of the following actions would be appropriate for Zhou to take? ( )

A. Tell Li's wife directly about his HIV status and ask her to monitor his condition

B. Help Li process his emotions, reduce his stress, and face his situation more positively

C. With Li's permission, connect him with an HIV support group

D. Present Li's full case details at the agency's case review meeting

E. Discuss strategies with Li's wife for improving their marriage

These questions are representative because they:
1. Test understanding of core social work concepts (intervention and ethics)
2. Present realistic scenarios social workers might encounter
3. Require application of theoretical knowledge to practical situations
4. Test both factual knowledge and professional judgment

# Appendix B: Model Prompts

Note: All testing procedures used the Chinese version of the prompt. English translations are provided for the reader's convenience.

**System Prompt For All Conditions**
- Chinese
- English

**Condition 1, Single-Select Questions - Chinese**

你是一位精通中国大陆社会政策和社会工作领域的专家，你正在参加中国社会工作者职业水平考试。你只能以指定的JSON格式回答，不能有任何其他对话或说明。

这是一道单选题。请按照以下方法分析题目：
1. 系统性评估每个选项
2. 使用排除法识别错误选项，选择一个最佳答案
3. 验证最终选择
4. 输出以下JSON格式并确保仅此格式：
{{
 "答案": "仅填写一个字母，如B",
 "理由": "用150字左右简明解释原因",
 "信心":"请用0-100之间的一个数字表示你对所给出答案的确定程度。"
}}
考题：{question}
选项：{selections}

**Condition 1, Single-Select Questions - English Translation**

You are an expert specializing in mainland China's social policies and social work, and you are taking the Chinese National Social Work Examination. You must respond only in the specified JSON format, with no other dialogue or explanations.
This is a multiple-choice, single-select question. Please analyze it using the following method:
1. Systematically evaluate each option
2. Use the process of elimination to identify incorrect options and select the best answer
3. Verify the final selection
4. Output in the following JSON format and ensure only this format:
{{
   "Answer": "Fill in only one letter, like B",
   "Reason": "Briefly explain the reason in about 150 words",
   "Confidence": "Use a number between 0-100 to indicate your certainty level for the given answer"
}}
Question: {question}
Options: {selections}

**Condition 1, SATA Questions - Chinese**

你是一位精通中国大陆社会政策和社会工作领域的专家，你正在参加中国社会工作者职业水平考试。你只能以指定的JSON格式回答，不能有任何其他对话或说明。

这是一道多选题，备选项中有两个或两个以上符合题意，至少有一个选项是错误的。请按照以下方法分析题目：
1. 独立评估每个选项，并仅保留最为确定的选项
2. 考虑选项之间的关系，但避免包含不完全确定的选项
3. 将所有选项作为整体进行验证
4. 输出以下JSON格式并确保仅此格式：
{{
  "答案": "仅列出你最有把握的选项的字母，如 A, C",
  "理由": "用150字左右简明解释原因",
  "信心":"请用0-100之间的一个数字表示你对所给出答案的确定程度。"
}}
考题：{question}
选项：{selections}

**Condition 1, SATA Questions - English Translation**

You are an expert specializing in mainland China's social policies and social work, and you are taking the Chinese National Social Work Examination. You must respond only in the specified JSON format, with no other dialogue or explanations.
This is a multiple-choice question where two or more options are correct, and at least one option is incorrect. Please analyze it using the following method:
1. Evaluate each option independently and only keep the most certain options
2. Consider relationships between options, but avoid including options that aren't completely certain
3. Verify all options as a whole
4. Output only in this JSON format with nothing else:
{{
   "Answer": "Only list the letters of options you're most confident about, e.g. A, C",
   "Reason": "Briefly explain your reasoning in about 150 words",
   "Confidence": "Express your certainty in the answer with a number between 0-100"
}}
Question: {question}
Options: {selections}

**Condition 2, Single-Select Questions - Chinese**

你是一位精通中国大陆社会政策和社会工作领域的专家，你正在参加中国社会工作者职业水平考试。你只能以指定的JSON格式回答，不能有任何其他对话或说明。
这是一道单选题。如果你不确定正确选项是哪个，请跳过该问题不作答。禁止猜测答案。如果你选择作答，请按照以下结构化方法分析题目：
1. 系统性评估每个选项
2. 使用排除法识别错误选项，选择一个最佳答案
3. 验证最终选择
4. 输出以下JSON格式并确保仅此格式：
{{
  "答案": "仅填写一个字母，如 B。如果你选择不作答，请填写"跳过"",
  "理由": "用150字左右简明解释原因。如果你选择不作答，请填写"跳过""
  "信心":"请用0-100之间的一个数字表示你对所给出答案的确定程度。如果你选择不作答，请填写"跳过""

}}
考题：{question}
选项：{selections}

**Condition 2, Single-Select Questions - English Translation**

You are an expert specializing in mainland China's social policies and social work, and you are taking the Chinese National Social Work Examination. You can only answer in the specified JSON format, with no other dialogue or explanations.

This is a multiple-choice, single-select question. If you are unsure which option is correct, please skip the question without answering. Refrain from guessing the answer. If you choose to answer, please analyze the question using the following structured method:
1. Systematically evaluate each option
2. Use the process of elimination to identify incorrect options and select the best answer
3. Verify the final selection
4. Output in the following JSON format and ensure only this format:
{{
    "Answer": "Fill in only one letter, like B. If you choose to skip, write 'skip'",
    "Reason": "Briefly explain the reason in about 150 words. If you choose to skip, write 'skip'",
    "Confidence": "Use a number between 0-100 to indicate your certainty level for the given answer. If you choose to skip, write 'skip'"
}}
Question: {question}
Options: {selections}

**Condition 2, SATA Questions - Chinese**

你是一位精通中国大陆社会政策和社会工作领域的专家，你正在参加中国社会工作者职业水平考试。你只能以指定的JSON格式回答，不能有任何其他对话或说明。
这是一道多选题，备选项中有两个或两个以上符合题意，至少有一个选项是错误的。如果你不确定正确选项是哪些，请跳过该问题不作答。禁止猜测答案。如果你选择作答，请按照以下结构化方法分析题目：
1. 独立评估每个选项，并仅保留最为确定的选项
2. 考虑选项之间的关系，但避免包含不完全确定的选项
3. 将所有选项作为整体进行验证
4. 输出以下JSON格式并确保仅此格式：
{{
    "答案"："仅列出你最有把握的选项的字母，如 A, C, D。如果你选择不作答，请填写"跳过"",
    "理由"："用150字左右简明解释原因。如果你选择不作答，请填写"跳过"" 
    "信心"："请用0-100之间的一个数字表示你对所给出答案的确定程度。如果你选择不作答，请填写"跳过""
}}
考题：{question}
选项：{selections}

**Condition 2, SATA Questions - English Translation**

You are an expert specializing in mainland China's social policies and social work, and you are taking the Chinese National Social Work Examination. You can only answer in the specified JSON format, with no other dialogue or explanations.
This is a multiple-choice question where two or more options are correct, and at least one option is incorrect. If you are unsure which options are correct, please skip the question without answering. Refrain from guessing the answer. If you choose to answer, please analyze the question using the following structured method:
1. Evaluate each option independently and only keep the most certain options
2. Consider relationships between options, but avoid including options that aren't completely certain
3. Verify all options as a whole
4. Output only in this JSON format with nothing else:
{{
    "Answer": "Only list the letters of options you're most confident about, e.g. A, C. If you choose to skip, write 'skip'",
    "Reason": "Briefly explain your reasoning in about 150 words. If you choose to skip, write 'skip'",
    "Confidence": "Express your certainty in the answer with a number between 0-100. If you choose to skip, write 'skip'"
}}
Question: {question}
Options: {selections}

**Condition 3, Single-Select Questions - Chinese**

你是一位精通中国大陆社会政策和社会工作领域的专家，你正在参加中国社会工作者职业水平考试。你只能以指定的JSON格式回答，不能有任何其他对话或说明。
你正在参加一个特殊版本的中国社会工作者职业水平考试。本题为单选题，只有选项，没有题干。虽然没有题干和上下文很难做出选择，但是你必须选择一个最佳答案，且在作答后描述你的决策过程。
输出以下JSON格式并确保仅此格式：
{{
  "答案": "仅填写一个字母，如 B",
  "理由": "简要解释你的决策过程"
}}
选项：{selections}

**Condition 3, Single-select Questions - English translation**

You are an expert specializing in mainland China's social policies and social work, and you are taking the Chinese National Social Work Examination. You can only answer in the specified JSON format, with no other dialogue or explanations.

You are taking a special version of the Chinese National Social Work Examination. This test consists of multiple-choice, single-select questions with only options provided, without the question stem. Although it's challenging to make a choice without the question stem and context, you must select the best answer and describe your decision-making process afterward. Output in the following JSON format and ensure only this format:
{{
    "Answer": "Fill in only one letter, like B",
    "Reason": "Briefly explain your decision-making process"

}}
Options: {selections}

**Condition 3, SATA Questions - Chinese**

你是一位精通中国大陆社会政策和社会工作领域的专家，你正在参加中国社会工作者职业水平考试。你只能以指定的JSON格式回答，不能有任何其他对话或说明。

你正在参加一个特殊版本的中国社会工作者职业水平考试。本题为多选题，备选项中有两个或两个以上符合题意，至少有一个选项是错误的。本题只有选项，没有题干。虽然没有题干和上下文很难做出选择，但是你必须做出选择，且在作答后描述你的决策过程。

输出以下JSON格式并确保仅此格式：

{{
    "答案": "列出所有正确选项的字母，如 A, C, D",
    "理由": "简要解释你的决策过程"
}}
选项: {selections}

**Condition 3, SATA Questions - English translation**

You are an expert specializing in mainland China's social policies and social work, and you are taking the Chinese National Social Work Examination. You can only answer in the specified JSON format, with no other dialogue or explanations.

You are taking a special version of the Chinese National Social Work Examination. This is a multiple-choice question where two or more options are correct, and at least one option is incorrect. This question consists solely of options and does not include the question stem. Although it's challenging to choose without the question stem and context, you must select the best answer and describe your decision-making process afterward. Output in the following JSON format and ensure only this format:

{{
    "Answer": "List all the letters of the correct options, like A, C, D",
    "Reason": "Briefly explain your decision-making process"
}}
Options: {selections}